\documentstyle[12pt]{article}                    %%%
\newfont{\ffont}{msym10}                        %%
\newcommand{\beq}{\begin{equation}}             %%
\newcommand{\eeq}{\end{equation}}               %%
\newcommand{\bqry}{\begin{eqnarray}}            %%
\newcommand{\eqry}{\end{eqnarray}}              %%
\newcommand{\bqryn}{\begin{eqnarray*}}          %%
\newcommand{\eqryn}{\end{eqnarray*}}            %%
\newcommand{\preprint}[1]{\begin{table}[t]      %%
            \begin{flushright}                  %%
            \begin{large}{#1}\end{large}        %%
            \end{flushright}                    %%
            \end{table}}                        %%
\newcommand{\PD}[2]                             %%
    {\frac{\partial^{#2}}{\partial #1^{#2}}}    %%
\begin{document}
\preprint{LA-UR-96-2499 \\ IASSNS-HEP-96/76}
\title{Mass Spectrum of a Meson Nonet is Linear}
\author{\\ L. Burakovsky\thanks{Bitnet: BURAKOV@QCD.LANL.GOV} \
\\  \\  Theoretical Division, T-8 \\  Los Alamos National  
Laboratory \\ Los
Alamos NM 87545, USA \\  \\  and  \\  \\
L.P. Horwitz\thanks{Bitnet: HORWITZ@SNS.IAS.EDU. On sabbatical leave from
School of Physics and Astronomy, Tel Aviv University, Ramat Aviv, Israel.
Also at Department of Physics, Bar-Ilan University, Ramat-Gan,  
Israel  } \
\\  \\ School of Natural Sciences \\ Institute for Advanced Study  
\\ Princeton
NJ 08540, USA \\}
%  \\ and \\  \\ W.C. Schieve\thanks{Bitnet: WCS@MAIL.UTEXAS.EDU}\
%\\  \\ Ilya Prigogine Center \\ for Studies in Statistical Mechanics \\
%University of Texas at Austin \\ Austin TX 78712, USA \\}
\date{ }
\maketitle
\begin{abstract}
It is argued that the mass spectrum of a meson nonet is linear,  
consistent
with the standard Gell-Mann--Okubo mass formula and leading to an extra 
Gell-Mann--Okubo mass relation for the masses of the isoscalar  
states. This
relation is shown to hold with an accuracy of up to $\sim $3\% for all
well-established nonets. It also suggests a new $q\bar{q}$  
assignment for the
scalar meson nonet.
\end{abstract}

\bigskip
\section*{  }
The hadronic mass spectrum is an essential ingredient in theoretical 
investigations of the physics of strong interactions. It is well  
known that
the correct thermodynamic description of hot hadronic matter requires 
consideration of higher mass excited states, the resonances, whose
contribution becomes essential at temperatures $\sim O(100$ MeV)
\cite{Shu,Leut}. The method for taking into account these
resonances was suggested by Belenky and Landau \cite{BL} as considering 
unstable particles on an equal footing with the stable ones in the
thermodynamic quantities; e.g., the formulas for the pressure and energy 
density in a resonance gas read\footnote{For simplicity, we neglect the 
chemical potential and approximate the particle statistics by the
Maxwell-Boltzmann one.}
\beq
p=\sum _ip_i=\sum _ig_i\frac{m_i^2T^2}{2\pi  
^2}K_2\left(\frac{m_i}{T}\right),
\eeq
\beq
\rho =\sum _i\rho _i,\;\;\;\rho _i=T\frac{dp_i}{dT}-p_i,
\eeq
where $g_i$ are the corresponding degeneracies ($J$ and $I$ are spin and
isospin, respectively), $$g_i=\frac{\pi ^4}{90}\times \left[
\begin{array}{ll}
(2J_i+1)(2I_i+1) & {\rm for\;non-strange\;mesons} \\
4(2J_i+1) & {\rm for\;strange}\;(K)\;{\rm mesons} \\
2(2J_i+1)(2I_i+1)\times 7/8 & {\rm for\;baryons}
\end{array} \right. $$
These expressions may be rewritten with the help of a {\it  
resonance spectrum,}
\beq
p=\int _{m_1}^{m_2}dm\;\tau (m)p(m),\;\;\;p(m)\equiv  
\frac{m^2T^2}{2\pi ^2}
K_2\left(\frac{m}{T}\right),
\eeq
\beq
\rho =\int _{m_1}^{m_2}dm\;\tau (m)\rho (m),\;\;\;\rho (m)\equiv
T\frac{dp(m)}{dT}-p(m),
\eeq
normalized as
\beq
\int _{m_1}^{m_2}dm\;\tau (m)=\sum _ig_i,
\eeq
where $m_1$ and $m_2$ are the masses of the lightest and heaviest  
species,
respectively, entering the formulas (1),(2).

In both the statistical bootstrap model \cite{Hag,Fra} and the dual  
resonance
model \cite{FV}, a resonance spectrum takes on the form
\beq
\tau (m)\sim m^a\;e^{m/T_0},
\eeq
where $a$ and $T_0$ are constants. The treatment of a hadronic  
resonance gas
by means of the spectrum (6) leads to a singularity in the thermodynamic 
functions at $T=T_0$ \cite{Hag,Fra} and, in particular, to an  
infinite number
of the effective degrees of freedom in the hadron phase, thus hindering a
transition to the quark-gluon phase. Moreover, as shown by Fowler  
and Weiner
\cite{FW}, an exponential mass spectrum of the form (6) is  
incompatible with
the existence of the quark-gluon phase: in order that a phase  
transition from
the hadron phase to the quark-gluon phase be possible, the hadronic  
spectrum
cannot grow with $m$ faster than a power.

In our previous work \cite{spectrum} we considered a model for a  
transition
from a phase of strongly interacting hadron constituents, described by a 
manifestly covariant relativistic statistical mechanics which  
turned out to be
a reliable framework in the description of realistic physical systems 
\cite{mancov}, to the hadron phase described by a resonance  
spectrum, Eqs.
(3),(4). An example of such a transition may be a relativistic high  
temperature
Bose-Einstein condensation studied by the authors in ref.  
\cite{cond}, which
corresponds, in the way suggested by Haber and Weldon \cite{HW}, to 
spontaneous flavor symmetry breakdown, $SU(3)_F\rightarrow SU(2)_I\times 
U(1)_Y,$ upon which hadronic multiplets are formed, with the masses  
obeying
the Gell-Mann--Okubo formulas \cite{GMO}
\beq
m^\ell =a+bY+c\left[ \frac{Y^2}{4}-I(I+1)\right];
\eeq
here $I$ and $Y$ are the isospin and hypercharge, respectively,  
$\ell $ is 2
for mesons and 1 for baryons, and $a,b,c$ are independent of $I$  
and $Y$ but,
in general, depend on $(p,q),$ where $(p,q)$ is any irreducible  
representation
of $SU(3).$ Then only the assumption on the overall degeneracy  
being conserved
during the transition is required to lead to the unique form of a  
resonance
spectrum in the hadron phase:
\beq
\tau (m)=Cm,\;\;\;C={\rm const}.
\eeq
Zhirov and Shuryak \cite{ZS} have found the same result on  
phenomenological
grounds. As shown in ref. \cite{ZS}, the spectrum (8), used in the  
formulas (3),(4) (with the upper limit of integration infinity), leads to
the equation of state $p,\rho \sim T^6,$ $p=\rho /5,$ called by  
Shuryak the
``realistic'' equation of state for hot hadronic matter \cite{Shu},  
which has
some experimental support. Zhirov and Shuryak \cite{ZS} have calculated 
the velocity of sound, $c_s^2\equiv dp/d\rho =c_s^2(T),$ with $p$  
and $\rho $
defined in Eqs. (1),(2), and found that $c_s^2(T)$ at first  
increases with $T$
very quickly and then saturates at the value of $c_s^2\simeq 1/3$  
if only the
pions are taken into account, and at $c_s^2\simeq 1/5$ if  
resonances up to
$M\sim 1.7$ GeV are included.

We have checked the coincidence of the results given by the linear  
spectrum (8)
with those obtained directly from Eq. (1) for the actual hadronic  
species with
the corresponding degeneracies, for all well-established multiplets, \\  
the mesons:

1 $^3S_1$ $J^{PC}=1^{--}$ nonet, $\;\rho (770),\;$ $\;\omega (783),\;$ 
$\;\phi (1020),\;$ $\;K^\ast (892)\;$

1 $^1P_1$ $J^{PC}=1^{+-}$ nonet, $b_1(1235),$ $h_1(1170),$ $h_1(1380),$ 
$K_1(1270)$

1 $^3P_1$ $J^{PC}=1^{++}$ nonet, $a_1(1260),$ $f_1(1285),$ $f_1(1510),$ 
$K_1(1400)$

1 $^3P_2$ $J^{PC}=2^{++}$ nonet, $a_2(1320),$ $f_2(1270),$  
$f_2^{'}(1525),$
$K_2^\ast (1430)$

1 $^3D_3$ $J^{PC}=3^{--}$ nonet, $\rho _3(1690),$ $\omega
_3(1670),$ $\phi _3(1850),$ $K^\ast _3(1780),$ \\ the baryons:

$J^P=\frac{1}{2}^{+}$ octet, $\;N(939),$ $\Lambda (1116),$ $\Sigma (
1190),$ $\Xi (1320)$

$J^P=\frac{3}{2}^{+}$ decuplet, $\Delta (1232),$ $\Sigma (
1385),$ $\Xi (1530),$ $\Omega (1672)$

$J^P=\frac{3}{2}^{-}$ nonet, $N(1520),$
$\Lambda (1690),$ $\Sigma (1670),$ $\Xi (1820),$ $\Lambda (1520)$

$J^P=\frac{5}{2}^{+}$ octet, $N(1680),$ $\Lambda (1820),$ $\Sigma  
(1915),$
$\Xi (2030),$ \\ and found it excellent \cite{spectrum}. Shown are  
typical
figures of ref. \cite{spectrum} in which the results given by both,  
Eq. (1),
and Eq. (3) with a linear spectrum, are compared.\footnote{Instead  
of a direct
comparison of Eqs. (1) and (3), we compared the expressions  
$p/p_{SB}$ for
both cases, where $p_{SB}\equiv \sum _ig_i\pi ^2/90\;T^4,$ i.e.,  
$p_{SB}$ is
the pressure in an ultrarelativistic gas with $g=\sum _ig_i$ degrees of 
freedom.} Thus, the theoretical implication that a linear spectrum  
is the
actual spectrum in the description of individual hadronic multiplets, is 
consistent with experiment as well. In our recent paper  
\cite{enigmas} we have
applied a linear spectrum to the problem of establishing the  
correct $q\bar{q}$
assignment for the problematic meson nonets, like the scalar,  
axial-vector and
tensor ones, and separating out non-$q\bar{q}$ mesons.

The easiest way to see that a linear spectrum corresponds to the actual 
spectrum of a meson nonet is as follows\footnote{For a baryon  
multiplet, it is
more difficult to show that the mass spectrum is linear, since the
Gell-Mann--Okubo formulas are linear in mass for baryons. More detailed 
discussion is given in \cite{spectrum}.}. Let us calculate the  
average mass
squared for a spin-$s$ nonet:
\beq
\langle m^2\rangle _9\equiv \frac{\sum  
_ig_i\;m_i^2}{\sum_ig_i}=\frac{3m_1^2+
4m_{1/2}^2+m_{0^{'}}^2+m_{0^{''}}^2}{9},
\eeq
where $m_1,\;m_{1/2},\;m_0,\;m_{0^{'}}$ are the masses of  
isovector, isospinor,
and two isoscalar states, respectively, and the spin degeneracy, $2s+1,$ 
cancels out. In general, the isoscalar states\footnote{The $\omega  
_{0^{'}}$
is a mostly octet isoscalar.} $\omega _{0^{'}}$ and $\omega  
_{0^{''}}$ are the
octet $\omega _8$ and singlet $\omega _0$ mixed states because of  
$SU(3)$
breaking, $$\omega _{0^{'}}=\omega _8\cos \theta _M- \omega _0\sin  
\theta _M,$$
$$\omega _{0^{''}}=\omega _8\sin \theta _M+ \omega _0\cos \theta  
_M,$$ where
$\theta _M$ is a mixing angle. Assuming that the matrix element of the 
Hamiltonian between the states yields a mass squared, i.e., $m_{0^{'}}^2=
\langle \omega _{0^{'}}|H|\omega _{0^{'}}\rangle $ etc., one  
obtains from the
above relations \cite{Per},
\beq
m_{0^{'}}^2=m_8^2\cos ^2\theta _M+m_0^2\sin ^2\theta  
_M-2m_{08}^2\sin \theta _M
\cos \theta _M,
\eeq
\beq
m_{0^{''}}^2=m_8^2\sin ^2\theta _M+m_0^2\cos ^2\theta  
_M+2m_{08}^2\sin \theta
_M\cos \theta _M.
\eeq
Since $\omega _{0^{'}}$ and $\omega _{0^{''}}$ are orthogonal, one  
has further
\beq
m_{0^{'}0^{''}}^2=0=(m_8^2-m_0^2)\sin \theta _M\cos \theta _M+  
m_{08}^2(\cos ^2
\theta _M-\sin ^2\theta _M).
\eeq
Eliminating $m_0$ and $m_{08}$ from (10)-(12) yields
\beq
\tan ^2\theta _M=\frac{m_8^2-m_{0^{'}}^2}{m_{0^{''}}^2-m_8^2}.
\eeq
It also follows from (10),(11) that, independent of $\theta _M,$  
$m_{0^{'}}^2+
m_{0^{''}}^2=m_8^2+m_0^2,$ and therefore, Eq. (9) may be rewritten as    
\beq
\langle m^2\rangle _9=\frac{3m_1^2+4m_{1/2}^2+m_8^2+m_0^2}{9}.
\eeq
For the octet, $(3\;m_1,\;4\;m_{1/2},\;1\;m_8),$ the  
Gell-Mann--Okubo formula
(as follows from (7)) is
\beq
4m_{1/2}^2=3m_8^2+m_1^2.
\eeq
Therefore, the average mass squared for the octet is
\beq
\langle m^2\rangle  
_8=\frac{3m_1^2+4m_{1/2}^2+m_8^2}{8}=\frac{m_1^2+m_8^2}{2},
\eeq
where Eq. (15) was used. In the exact $SU(3)$ limit where the $u,$  
$d$ and $s$
quarks have equal masses, all the squared masses of the nonet  
states are equal
as well. Since in this limit all the squared masses of the octet  
states are
equal to the average mass squared of the octet\footnote{In a manifestly 
covariant theory, this holds since a total mass squared is rigorously 
conserved. In the standard framework, for pseudoscalar mesons, this  
is easily
seen by using the lowest order relations \cite{GMOR} $m_1^2\equiv  
m_\pi ^2=
2mB,$ $m_{1/2}^2\equiv m_K^2=(m+m_s)B,$ where $m=(m_u+m_d)/2,$ and  
$B$ is
related to the quark condensate. Therefore, it follows from  
(15),(16) that
$m_8^2=2/3\;(2m_s+m)B,$ $\langle m^2\rangle  
_8=2/3\;(m_s+2m)B=2/3\;(m_u+m_d+
m_s)B.$ In the exact $SU(3)$ limit, $m_u=m_d=m_s=\bar{m},$ and  
hence $m_1^2=
m_{1/2}^2=m_8^2=\langle m^2\rangle _8=2\bar{m}B.$ For higher mass  
mesons,
since the states with equal isospin (and alternating parity) lie on  
linear
Regge trajectories, one may expect the relations of the form  
$(c=C/B)$ $m_1^2=
2mB+C=(2m+c)B,$ $m_{1/2}^2=(m+m_s)B+C=(m+m_s+c)B,$  
$m_8^2=2/3\;(2m_s+m)B+C=
2/3\;(2m_s+m+3/2\;c)B,$ consistent with the Gell-Mann--Okubo  
formula (15),
leading to $m_1^2=m_{1/2}^2=m_8^2=\langle m^2\rangle  
_8=2\bar{m}B+C$ in the
$SU(3)$ limit $m_u=m_d=m_s=\bar{m}.$ For vector mesons, such a  
relation was
obtained by Bal\'{a}zs in the flux-tube fragmentation approach to a  
low-mass
hadronic spectrum \cite{Bal}, $m_\rho ^2=m_\pi ^2+1/2\alpha ^{'},$  
with $\alpha
^{'}$ being a universal Regge slope, in good agreement with the  
experiment.},
Eq. (16), the mass of the singlet should have the same  
value,\footnote{It is
also seen from the relations of a previous footnote: since the total mass
squared of a nonet is proportional to the total mass of quarks the nonet 
members are made of, $\sum _ig_i\;m_i^2=(12m+6m_s)B+9C,$ it follows  
from the
above expressions for $m_1^2,$ $m_{1/2}^2$ and $m_8^2$ that  
$m_0^2=2/3\;(2m+m_
s)B+C=\langle m^2\rangle _8=\langle m^2\rangle _9.$} i.e.,
\beq
m_0^2=\frac{m_1^2+m_8^2}{2}.
\eeq
With Eq. (15), it then follows from (17) that $$m_0^2+m_8^2=2m_{1/2}^2,$$
which reduces, through $m_0^2+m_8^2=m_{0^{'}}^2+m_{0^{''}}^2,$ to
\beq
m_{0^{'}}^2+m_{0^{''}}^2=2m_{1/2}^2,
\eeq
which is an extra Gell-Mann--Okubo mass relation for a nonet. We  
will check
this relation below, and show that with the experimentally  
available meson
masses, the relative error in the values on the l.h.s. and r.h.s.  
of Eq. (18)
does not exceed 3\% for all well-established nonets (except for the 
pseudoscalar nonet for which Eq. (18) does not hold, perhaps  
because the $\eta
_0$ develops a large dynamical mass due to axial $U(1)$ symmetry  
breakdown
before it mixes with the $\eta _8$ to form the physical $\eta $ and  
$\eta ^{'}$
states). For a singlet-octet mixing close to ``ideal'' one, $\tan  
\theta _M
\simeq 1/\sqrt{2};$ it then follows from (13) that  
$$2m_{0^{'}}^2+m_{0^{''}}^
2\simeq 3m_8^2,$$ which reduces, through (15),(18), to
\beq
m_{0^{''}}\simeq m_1.
\eeq
Now it follows clearly why the ground states of all well-established 
nonets\footnote{This is also true for $q\bar{q}$ assignment of the  
scalar
meson nonet suggested by the authors in ref. \cite{enigmas}.}  
(except for the
pseudoscalar one) are almost mass degenerate pairs, like $(\rho
,\omega ).$\footnote{It follows from the relations of footnote 5  
that, in the
close-to-ideal mixing case, $m_{0^{'}}^2\simeq 2m_sB+C$ and $m_{0^{''}}^
2\simeq 2mB+C=m_1^2.$} In the close-to-ideal mixing case, Eq. (18)  
may be
rewritten, with the help of (19), as
\beq
m_1^2+m_{0^{'}}^2\simeq 2m_{1/2}^2.
\eeq
This relation for pseudoscalar and vector mesons with the ground  
states being
the mass degenerate pairs $(\pi ,\eta _0)$ and $(\rho ,\omega ),$  
respectively,
was previously obtained by Bal\'{a}zs and Nicolescu using the
dual-topological-unitarization approach to the confinement region  
of hadronic
physics (Eq. (21) of ref. \cite{BN}). With (16) and (17), Eq. (10)  
finally
reduces to
\beq
\langle m^2\rangle _9=\frac{m_1^2+m_8^2}{2},
\eeq
which, of course, coincides with both, $\langle m^2\rangle _8$ in  
(16) and
$m_0^2$ in (17), which is the property of the $SU(3)$ limit (or the 
conservation of a total mass squared in a manifestly covariant theory). 

For the actual mass spectrum of the nonet, the average mass squared  
(9) may be
represented in the form\footnote{Since $m_s>m,$ $m_1<m_{1/2}<m_8,$  
as seen in
the relations of footnote 5. Moreover, $m_1<m_0<m_8,$ and therefore, 
the range of integration in Eq. (22) is $(m_1,\;m_8).$}
\beq
\langle m^2\rangle _9=\frac{\int _{m_1}^{m_8}dm\;\tau (m)\;m^2}{\int
_{m_1}^{m_8}dm\;\tau (m)},
\eeq
and one sees that the only choice for $\tau (m)$ leading to the  
relation (21)
is $\tau (m)=Cm,$ $C={\rm const.}$ Indeed, in this case $$\langle  
m^2\rangle
_9=\frac{\int _{m_1}^{m_8}dm\;m^3}{\int  
_{m_1}^{m_8}dm\;m}=\frac{(m_8^4-m_1^
4)/4}{(m_8^2-m_1^2)/2}=\frac{m_1^2+m_8^2}{2},$$ in agreement with (21).

We now wish to check the formula (18) for all well-established  
meson nonets
(we indicate the actual particle masses, as given by the recent  
Particle Data
Group \cite{data}, and take $m_\pi =1/3\;(2m_{\pi ^{\pm }}+m_{\pi ^0}),$ 
$m_K=1/2\;(m_{K^{\pm }}+m_{K^0})$ etc.):

1) 1 $^1S_0$ $J^{PC}=0^{-+},$ $\pi (138),$ $\eta (547),$ $\eta  
^{'}(958),$
$K(495).$ In the assumption of no mixing of the $\eta _8$ and $\eta  
_0$ states,
it follows from (15),(17) that $$m_\eta ^2=m_{\eta  
_8}^2=1/3\;(4m_K^2-m_\pi ^
2)\simeq 4/3\;m_K^2,$$ $$m_{\eta ^{'}}^2=m_{\eta  
_0}^2=1/3\;(2m_K^2+m_\pi ^2)
\simeq 2/3\;m_K^2,$$ so that\footnote{These relations for $m_\eta $ and 
$m_{\eta ^{'}}$ are also contained in ref. [1], p. 20.} $$m_\eta \simeq 
\sqrt{4/3}\;m_K\approx 566\;{\rm MeV},$$ in fair agreement with the  
experiment,
but $$m_{\eta ^{'}}\simeq \sqrt{2/3}\;m_K\approx 400\;{\rm MeV},$$  
in strong
disagreement with the experiment. The reason for the invalidity of  
Eq. (18) for
the pseudoscalar nonet is, probably, a large dynamical mass of the  
$\eta _0$
due to axial $U(1)$ symmetry breakdown developed before it mixes  
with the
$\eta _8$ to form the physical $\eta $ and $\eta ^{'}$ states.

2) 1 $^3S_1$ $J^{PC}=1^{--},$ $\rho (769),$ $\omega (783),$ $\phi  
(1019),$
$K^\ast (894).$ In this case one has 1.65 GeV$^2$ on the l.h.s. of  
Eq. (18) vs.
1.60 GeV$^2$ on the r.h.s.

3) 1 $^1P_1$ $J^{PC}=1^{+-},$ $b_1(1231),$ $h_1(1170),$ $h_1(1380),$ 
$K_1(1273).$ Now one has 3.27 GeV$^2$ on the l.h.s. of (18) vs.  
3.24 GeV$^2$
on the r.h.s.

4) 1 $^3P_1$ $J^{PC}=1^{++},$ $a_1(1230),$ $f_1(1282),$ $f_1(1512),$
$K_1(1402).$ In this case the both values on the different sides of  
Eq. (18)
are equal to 3.93 GeV$^2.$

5) 1 $^3P_2$ $J^{PC}=2^{++},$ $a_2(1318),$ $f_2(1275),$ $f_2^{'}(1525),$
$K_2^\ast (1425).$ Now one has 3.95 GeV$^2$ on the l.h.s. of (18)  
vs. 4.06
GeV$^2$ on the r.h.s.

6) 1 $^3D_3$ $J^{PC}=3^{--},$ $\rho _3(1680),$ $\omega _3(1668),$  
$\phi _3(
1854),$ $K_3^\ast (1770).$ In this case one has 6.23 GeV$^2$ on the  
l.h.s. of
Eq. (18) vs. 6.27 GeV$^2$ on the r.h.s.

Thus, Eq. (18) holds with an accuracy of up to $\sim $3\% for the  
vector and
tensor meson nonets and with a higher accuracy for other nonets; for the 
axial-vector nonet it is exact.

7) Even for the 2 $^3P_2$ $J^{PC}=2^{++}$ nonet for which there is  
no isovector
candidate at present, and two of the three remaining states need  
experimental
confirmation\footnote{These are the $f_2(1810)$ and $K_2^\ast (1980).$},
$f_2(1811),$ $f_2(2011),$ $K_2^\ast (1975),$ one has 7.32 GeV$^2$  
on the l.h.s.
of (18) vs. 7.80 GeV$^2$ on the r.h.s., with a still satisfactory  
accuracy of
$\sim $6\%.

8) Let us now dwell upon a problematic nonet, the scalar meson one,  
whose
currently adopted $q\bar{q}$ asignment is \cite{data} $a_0(982),$  
$f_0(1300),$
$f_0(980),$ $K_0^\ast (1429).$ It is seen that for this assignment  
Eq. (18)
does not hold $(980<1300<1429).$ As we have shown, this relation  
should hold,
independent of a mixing angle (i.e., even for a ``non-ideal''  
nonet), and there
is no apparent physical reason for the mass shifts of the scalar  
nonet members,
like that of the $\eta _0,$ which would lead to the invalidity of  
Eq. (18).
Therefore, we conclude that the currently adopted $q\bar{q}$  
assignment of the
scalar meson nonet is $incorrect.$

There are five established isoscalars with $J^{PC}=0^{++},$ the  
$f_0(980),
f_0(1300),f_0(1370),$ $f_0(1525)$ and $f_0(1590).$ In the quark  
model, one
expects two 1 $^3P_0$ states and one 2 $^3P_0$  
$(u\bar{u}+d\bar{d})$-like
state below 1.8 GeV. Therefore, at least two of the five cannot  
find a place
in the quark model. Many broad $\pi \pi $ elastic resonances  
claimed in the
past (like the $\sigma (700),\;\epsilon (1200),\;f_0(1400))$ were  
collected by
the recent Particle Data Group under one entry, the $f_0(1300).$  
Similarly,
all the broad $\pi \pi $ inelastic $S$-wave resonance claims were  
collected
under one entry, the $f_0(1370),$ although they could be the $f_0(1300)$ 
provided the inelasticity of the latter is in fact larger than is  
presently
believed \cite{data1}. Although it is currently adopted as a member  
of the
nonet, there exists an interpretation of the $f_0(980)$ as a $K\bar{K}$ 
molecule \cite{Wei} since it [and the $a_0(980)]$ lies just below the 
$K\bar{K}$ threshold which is 992 MeV \cite{Flatte}. If the  
$f_0(980)$ is not
the 1 $^3P_0$ $s\bar{s}$ state, the latter should be found near  
1500 MeV with
partial decay widths close to the flavor symmetry predictions for  
an ideal
nonet \cite{Torn90}. The weak signal as 1515 MeV claimed by the  
LASS group
\cite{LASS} does not have the expected large width. In this case,
the $f_0(1525)$ (or $f_0(1520))$ could be a candidate for the 1 $^3P_0$ 
$s\bar{s}$ state \cite{Mont}. This $f_0(1525)$ has been identified as 
$K\bar{K}$ $S$-wave intensity peaking at the mass of the  
$f_2^{'}(1525)$ and
having a comparable width \cite{Aston,Baub}. The $f_0(1520)$ (as well as 
$f_0(1370))$ has been recently
observed by the Crystal Barrel Collaboration in a simultaneous fit  
to the
$\bar{p}p\rightarrow 3\pi ^0$ and $\bar{p}p\rightarrow \eta \eta  
\pi ^0$ data
\cite{Anis}. The both, $f_0(1525)$ and $f_0(1520)$ are adopted by  
the recent
Particle Data Group as one entry, the $f_0(1525).$ The $f_0(1590)$  
has been
seen in $\pi ^{-}p$ reactions at 38 GeV/c \cite{Binon,Alde87}. It has a 
peculiar decay pattern for $$\pi ^0\pi ^0:K\bar{K}:\eta \eta :\eta  
\eta ^{'}:4
\pi ^0=\;<0.3:\;<0.6:1:2.7:0.8,$$ which could favor a gluonium  
interpretation
\cite{Ger}. Another possibility is that it is a large deuteron-like  
$(\omega
\omega -\rho \rho )/\sqrt{2}$ bound state (``deuson'') \cite{Torn91}.

The established isovector with $J^{PC}=1^{++}$ is the $a_0(980).$  
Its mass,
$(982\pm 2)$ MeV, is low, compared to its isovector partners, like the 
$a_1(1260),$ $a_2(1320)$ and $b_1(1235).$ Its apparent width (as  
measured in
its $\eta \pi $ decay mode), $(54\pm 10)$ MeV, is small, compared to ist 
partners (which have 100 MeV and more). Moreover, neither the  
relative coupling
of the $a_0(980)$ to $\eta \pi $ and $K\bar{K},$ nor its width to  
$\gamma
\gamma ,$ are known well enough to draw firm conclusions on its nature 
$(q\bar{q}$, $2q2\bar{q}$ state, $K\bar{K}$ molecule, etc.). If the  
$a_0(980)$
is not the $^3P_0$ state, the latter should be observed near 1300  
MeV, with
partial decay width as expected from flavor symmetry. The candidate  
$a_0(1320)$
identified by GAMS as intensity peaking at the mass of the  
$a_2(1320)$ and
having a comparable width \cite{Bout}, needs experimental confirmation.

Thus, an attractive choice for the $q\bar{q}$ scalar meson nonet  
could be
the $a_0(1320),$ $K_0^\ast (1430),$ $f_0(1300)$ (or $f_0(1370))$ and 
$f_0(1525)$ (or $f_0(1520)).$ This choice would leave out the  
$a_0(980)$ and
$f_0(980)$ which could be then interpreted in terms of four-quark or 
$K\bar{K}$ molecule states, and one may then speculate that the  
$f_0(1590)$ is
a glueball, or, at least, a state rich in glue. For the $q\bar{q}$  
assignment
suggested above, one has $4.00-4.20$ GeV$^2$ on the l.h.s. of Eq.  
(18) (4.00
corresponds to the assignment which includes $f_0(1300)$ and  
$f_0(1520)$ while
4.20 to that with $f_0(1370)$ and $f_0(1525))$ vs. 4.08 GeV$^2$ on  
the r.h.s.,
and we conclude that for this assignment, the formula (18) holds,  
with a high
accuracy, as should be the case for a genuine meson nonet.

9) Similar analysis for the 1 $^3D_1$ $J^{PC}=1^{--}$ and 2 $^3S_1$  
$J^{PC}=
1^{--}$ nonets will be given in a separate publication \cite{separate}. 

Evidently, one may choose an opposite way, viz., starting from a linear 
spectrum as the actual spectrum of a nonet, to derive the  
Gell-Mann--Okubo mass
formula. To this end, one should first calculate the average mass  
squared, Eq.
(21). Then one has to place 9 nonet states in the interval  
$(m_1,\;m_8)$ in a
way that preserves the average mass squared. As we already know,  
the isoscalar
singlet mass squared should coincide with the average mass squared;  
for the
remaining 8 states one would have the relation (16) which would in  
turn reduce
to the Gell-Mann--Okubo formula (15). One sees that the assumption  
on a linear
mass spectrum turns out to be a good alternative to the group  
theoretical
mechanism of symmetry breaking, for the derivation of the  
Gell-Mann--Okubo type
relations, which may be rather difficult technical task for a  
higher symmetry
group.

The method of the derivation of the Gell-Mann--Okubo mass relations  
described
above may be easily generalized to the case of four or more  
flavors. In our
recent paper \cite{recent}, by applying this method to an $SU(4)$  
hexadecuplet,
we have derived the corresponding Gell-Mann--Okubo mass formula and  
found it
to be in good agreement with the experimentally established masses  
of the
charmed mesons.

\section*{Acknowledgements}
One of us (L.B.) wish to thank E.V. Shuryak for very valuable  
discussions on
hadronic resonance spectrum.

%\bigskip
%\bigskip

\newpage
\centerline{FIGURE CAPTIONS}
\bigskip
\bigskip
\bigskip
\bigskip
\hfil\break
Fig. 1. Temperature dependence of the ratio $p/p_{SB}$ as  
calculated from:
a) Eq. (1), b) Eq. (3) with a linear spectrum, for the 1 $^1P_1$ $J^{PC}=
1^{+-}$ meson nonet, $b_1(1235),$ $h_1(1170),$ $h_1(1380),$
$K_1(1270).$\hfil\break
\hfil\break
\hfil\break
\hfil\break
Fig. 2. The same as Fig. 1 for the 1 $^3P_2$ $J^{PC}=2^{++}$ meson  
nonet,
$a_2(1320),$ $f_2(1270),$ $f_2^{'}(1525),$ $K_2^\ast (1430).$\hfil\break
\hfil\break
\hfil\break
\hfil\break
Fig. 3. The same as Fig. 1 for the $J^P=\frac{1}{2}^{+}$ baryon octet,
$N(939),$ $\Lambda (1116),$ $\Sigma (1193),$ $\Xi (1318).$\hfil\break
\hfil\break
\hfil\break
\hfil\break
Fig. 4. The same as Fig. 1 for the $J^P=\frac{3}{2}^{-}$ baryon nonet,
$N(1520),$ $\Lambda (1690),$ $\Sigma (1670),$ $\Xi (1820),$
$\Lambda (1520).$\hfil\break
%\hfil\break
%\hfil\break
%\hfil\break
%Fig. 5. The same as Fig. 3 but with the $f_0(1240),$ $f_0(1400)$  
replaced by
%$f_0(1525),$ $f_0(1590).$\hfil\break
%\hfil\break
%\hfil\break
%\hfil\break
%Fig. 6. The same as Fig. 1 but for the scalar meson nonet with the  
assignment
%$a_0(1320),$ $f_0(1240),$ $f_0(1525),$ $K_0^\ast (1430).$\hfil\break
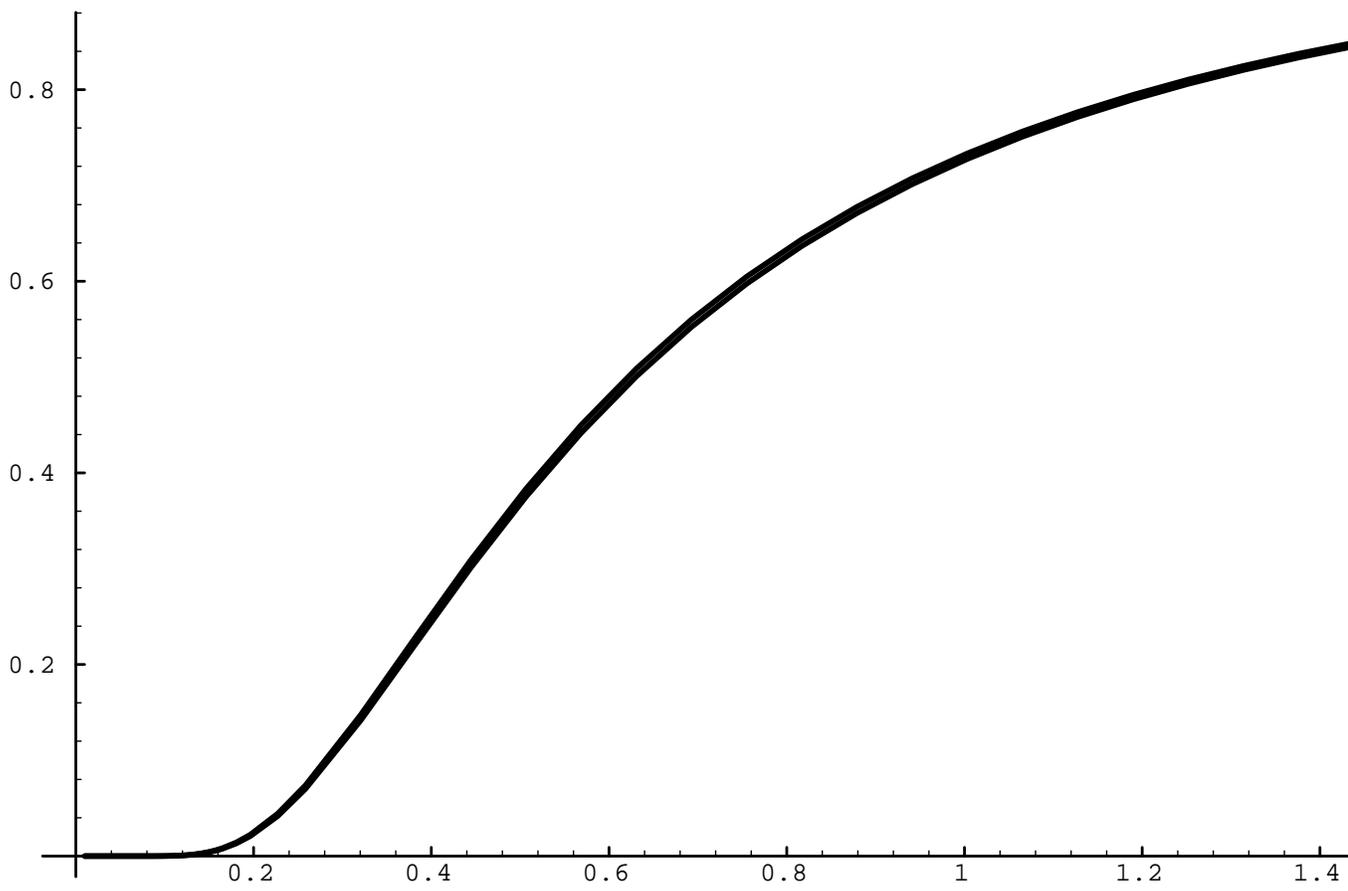
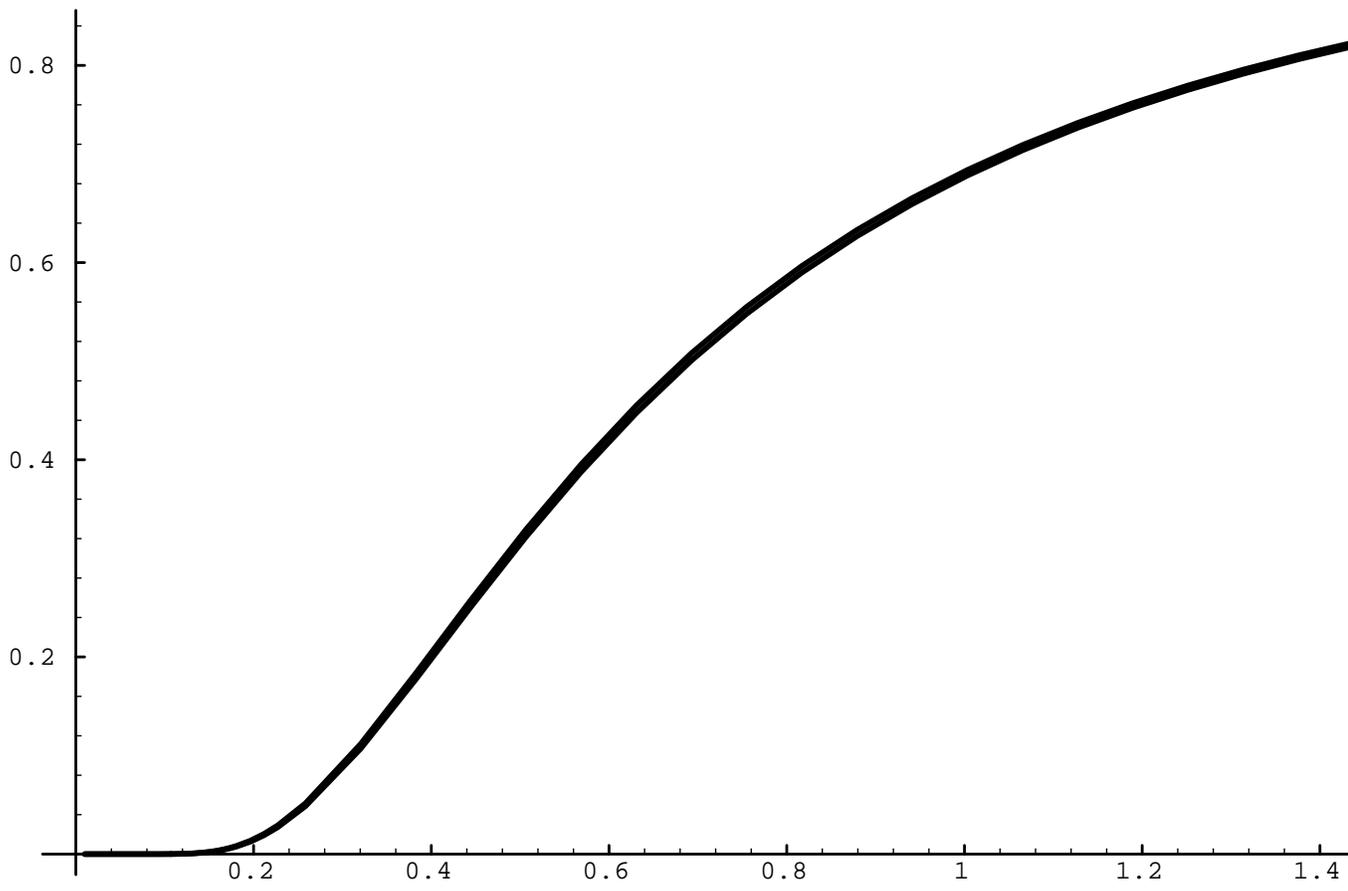
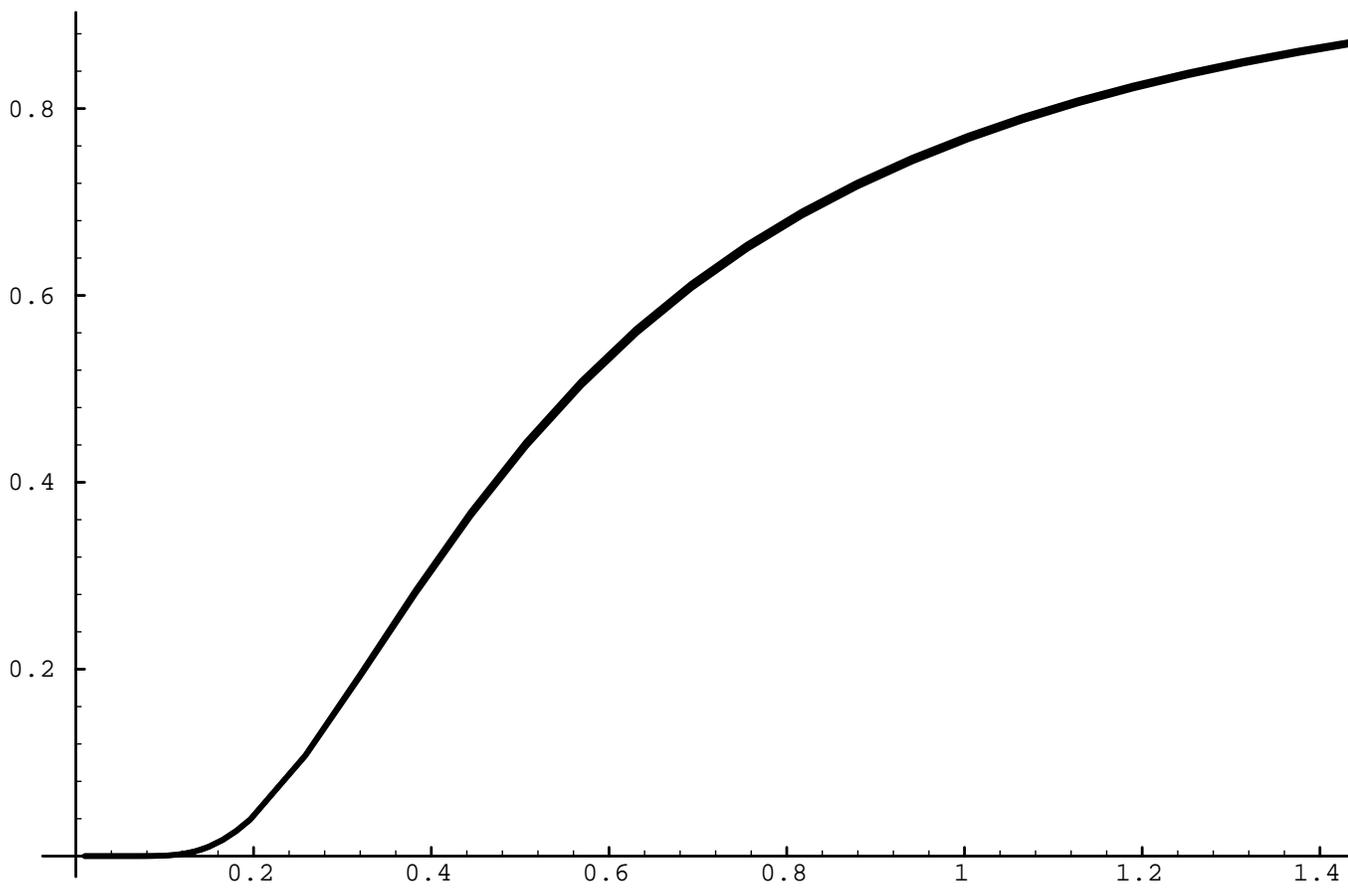
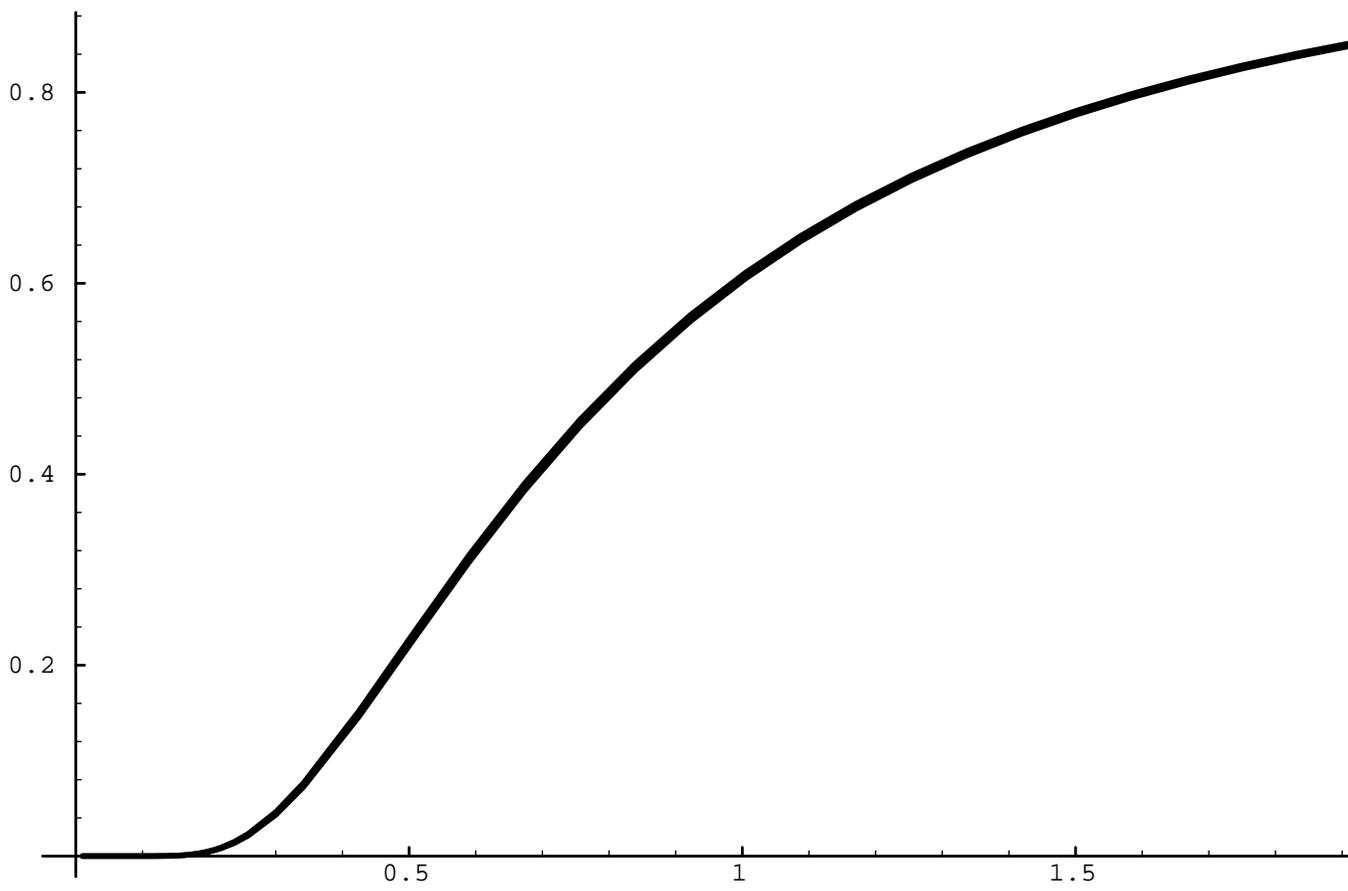
\end{document}